\documentclass[prd,twocolumn,groupedaddress,showpacs,nofootinbib]{revtex4}
\usepackage{graphicx}
\usepackage{amsmath}
\usepackage{epsfig}

\begin{document}

\title{Unified picture for Dirac neutrinos, dark matter, dark energy and matter-antimatter asymmetry}

\author{Pei-Hong Gu}

\affiliation{The Abdus Salam International Centre for Theoretical
Physics, Strada Costiera 11, 34014 Trieste, Italy}

\begin{abstract}

We propose a unified scenario to generate the masses of Dirac
neutrinos and cold dark matter at the TeV scale, understand the
origin of dark energy and explain the matter-antimatter asymmetry of
the universe. This model can lead to significant impact on the Higgs
searches at LHC.

\end{abstract}


\maketitle

Strong evidences from neutrino oscillation experiments have
confirmed the tiny neutrino masses of the order of
$10^{-2}\,\textrm{eV}$ \cite{pdg2006}. However, the neutrino's
Majorana or Dirac nature is still unknown. The smallness of the
neutrino masses can be elegantly understood via the Majorana
\cite{minkowski1977} or Dirac \cite{rw1983,gh2006} seesaw mechanism
in various extensions of the standard model (SM). The nature of the
dark matter, which contributes about $20\%$ \cite{pdg2006} to the
energy density of the unverse, also indicates the necessity of
supplementing to the existing theory. Currently many supersymmetric
or nonsupersymmetric candidates
\cite{blz2003,bmn2006,ma1977,sz1985,ss2007,eramo2007,gu2007} for the
dark matter have been proposed to study and search for. As for the
dark energy with the energy density $\sim(3\times
10^{-3}\,\textrm{eV})^4$ \cite{pdg2006}, which accelerates the
expansion of our universe at present, it is striking that its scale
is far lower than all the known scales in particle physics except
that of the neutrino masses. The intriguing coincidence between the
neutrino mass scale and the dark energy scale inspires us to
consider them in a unified scenario, as in the neutrino dark energy
model \cite{gwz2003}. The origin of the observed matter-antimatter
asymmetry \cite{pdg2006} of the universe poses another big challenge
to the SM, but within the Majorana or Dirac seesaw scenario, it can
be naturally explained through leptogenesis \cite{fy1986} or
neutrinogenesis \cite{dlrw1999}.

In this paper, we unify the mass origin for the Dirac neutrinos and
the dark matter in a nonsupersymmetric extension of the SM. After a
new $U(1)$ gauge symmetry is spontaneously broken at the TeV scale,
the SM neutrinos will obtain small Yukawa couplings to the new
right-handed neutrinos and the SM Higgs while other new introduced
fermions, which guarantee the theory free of gauge anomaly, will
acquire masses of a few hundred GeV. These new fermions with the
right amount of the relic density can serve as the candidate for the
cold dark matter. In order to understand the origin of the dark
energy, we further introduce a proper global symmetry, after which
is spontaneously broken near the Planck scale, a
pseudo-Nambu-Goldstone boson (pNGB) associated with the neutrino
mass-generation can explain the nature of the dark energy.
Meanwhile, the matter-antimatter asymmetry can be resolved via the
neutrinogenesis mechanism. This model predicts new Higss
phenomenology that can be tested at LHC.

\begin{figure} \vspace{4.5cm}
\epsfig{file=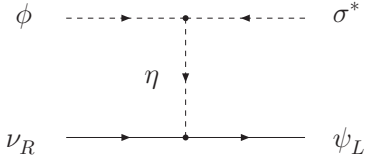, bbllx=6.0cm, bblly=7.0cm, bburx=16cm,
bbury=17cm, width=7.5cm, height=7.5cm, angle=0, clip=0}
\vspace{-9cm} \caption{\label{massgeneration} The dim-5 operator for
neutrino mass-generation. }
\end{figure}

To generate the masses for the Dirac neutrinos, we can simply
introduce three right-handed neutrinos to the SM. However, the
Yukawa couplings of the Dirac neutrinos should be extremely small.
One possibility to naturally explain this phenomena is to consider
the Dirac seesaw \cite{rw1983,gh2006}, in which the Yukawa couplings
of the neutrinos to the SM Higgs are generated by integrating out
some heavy particles, meanwhile, the conventional Yukawa couplings
of the neutrinos to the SM Higgs should be exactly forbidden by
consideration of symmetry. Here we consider a $U(1)_{X}$ gauge
symmetry, under which the right-handed neutrinos carry the charge
$-1$ while all SM particles transform trivially. We also introduce a
new $SU(2)_{L}^{}$ Higgs doublet which carries the same
$U(1)_{Y}^{}$ hypercharge with the SM Higgs but has the
$U(1)_{X}^{}$ charge $+1$. Thus the neutrinos can have the Yukawa
couplings to this new Higgs,
\begin{eqnarray}
\label{yukawa1} \mathcal{L} &\supset&
-\,y\,\overline{\psi_L^{}}\,\eta\, \nu_{R}^{}\,+\,\textrm{h.c.}\,,
\end{eqnarray}
where $\psi_{L}^{}$, $\eta$ and $\nu_{R}^{}$ are the SM lepton
doublets, new Higgs doublet and right-handed neutrinos,
respectively. However, different from the SM Higgs, the new one has
a positive quadratic term in the scalar potential so that it can not
develop a vacuum expectation value (\textit{vev}) to generate the
neutrino masses at this stage. Fortunately, we can conveniently
introduce a SM Higgs singlet with the $U(1)_{X}^{}$ charge $+1$ and
then obtain a trilinear coupling among three types of Higgs fields,
\begin{eqnarray}
\label{cubic} \mathcal{L} &\supset&
-\,\mu\,\sigma\,\eta^{\dagger}_{}\,\phi\,+\,\textrm{h.c.}\,,
\end{eqnarray}
where $\sigma$ and $\phi$ are the SM singlet and doublet Higgs
scalars, respectively. Therefore, as shown in Fig.
\ref{massgeneration}, we can obtain a dim-5 operator,
\begin{eqnarray}
\label{dimension5} \mathcal{L} &\supset&
\frac{\mu}{M_{\eta}^{2}}\,y\,\overline{\psi_{L}^{}}\,\phi\,\nu_{R}^{}\,\sigma\,+\,\textrm{h.c.}\,
\end{eqnarray}
by integrating out the new Higgs doublet. Once the $U(1)_{X}^{}$
symmetry is spontaneously broken by the \textit{vev},
$\langle\sigma\rangle$, the neutrinos will acquire the effective
Yukawa couplings to the SM Higgs. It is straightforward to see that
the effective Yukawa couplings can be highly suppressed by
$\mu\langle\sigma\rangle/M_{\eta}^{2}$ and hence the neutrinos will
obtain the tiny Dirac masses,
\begin{eqnarray}
\label{neutrinomass1}
m_{\nu}^{}&=&y_{eff}^{}\,\langle\phi\rangle\equiv\,-\,
\frac{\mu\,\langle\sigma\rangle}{M_{\eta}^{2}}\,y\,\langle\phi\rangle\,.
\end{eqnarray}
For instance, we find that by inputting
\begin{eqnarray}
\label{parameter} &\langle \phi\rangle\,\simeq\,
174\,\textrm{GeV},\,\,\langle\sigma\rangle\,=\,\mathcal{O}(\textrm{TeV})\,,&\nonumber\\
&\mu/M_{\eta}^{}\,=\,\mathcal{O}(0.1)\,,\,
M_{\eta}^{}\,=\,\mathcal{O}(10^{13-15}_{}\,\textrm{GeV})\,,&
\end{eqnarray}
the Yukawa couplings of the neutrinos can naturally remain small,
$y_{eff}^{}\sim \mathcal{O}(10^{-13}_{})$ for $y \sim
\mathcal{O}(10^{-2}_{}\,-\,1)$, and hence, the neutrino masses
become of the order of $m_{\nu}^{}\sim\mathcal{O}(10^{-2}_{}
\,\textrm{eV})$, which is consistent with the neutrino oscillation
data \cite{pdg2006}. In fact, as shown in \cite{gh2006}, by
minimizing the full scalar potential, the new Higgs doublet will
acquire a small \textit{vev},
\begin{eqnarray}
\label{veveta} \left<\eta\right> &\simeq& -\,\frac{\mu\,
\langle\sigma\rangle}{M_{\eta}^{2}}\,\langle\phi\rangle\,
\end{eqnarray}
with the range,
\begin{eqnarray}
\label{range}
10^{-2}_{}\,\textrm{eV}\,\lesssim\,\langle\eta\rangle\,\lesssim
\,1\,\textrm{eV}\,
\end{eqnarray}
for the parameters (\ref{parameter}). This confirms Eq.
(\ref{neutrinomass1}) due to the mass formula,
\begin{eqnarray}
\label{neutrinomass2} m_{\nu} &=& y\,\left<\eta\right>\,
\end{eqnarray}
from Eq. (\ref{yukawa1}).

The requirement to ensure anomaly free indicates the necessity of
supplementing the existing theory with three left-handed SM singlet
fermions $\chi_{L}^{}$ with the $U(1)_{X}^{}$ charge $+1$. Under the
present gauge symmetry, it is convenient to introduce three
right-handed singlet fermions $\chi_{R}^{}$ to generate the
following Yukawa couplings,
\begin{eqnarray}
\label{yukawa2} \mathcal{L} &\supset&
-\,f\,\sigma\,\overline{\chi_{L}^{}}\,\chi_{R}^{}\,+\,\textrm{h.c.}\,,
\end{eqnarray}
through which the singlet fermions will acquire masses,
\begin{eqnarray}
\label{darkmass1}m_{\chi}^{}&=&f\,\langle \sigma\rangle\,,
\end{eqnarray}
after the $U(1)_{X}^{}$ breaking by $\langle \sigma\rangle$. We
further consider a $Z_{3}^{}$ discrete symmetry, under which
$\chi_{L,R}^{}$ have the transformation properties
$\chi_{L,R}^{}\rightarrow \omega\, \chi_{L,R}^{}$ with
$\omega^{3}_{}=1$ while all other fields are trivial. In
consequence, the Yukawa coupling
$\overline{\psi_{L}^{}}\,\phi\,\chi_{R}^{}$ and the Majorana mass
term of $\chi_{R}^{}$ are exactly forbidden. So, $\chi_{L,R}^{}$
have not any decay modes and hence are inert. By diagonalizing the
mass matrix (\ref{darkmass1}), the inert fermions can be defined in
their mass-eigenbasis $\chi_{1,2,3}^{}$ with the corresponding
masses,
\begin{eqnarray}
\label{darkmass2} m_{\chi_{1,2,3}^{}}^{}&=&f_{1,2,3}^{}\,\langle
\sigma\rangle\,,
\end{eqnarray}
where $f_{1}^{}\leq f_{2}^{}\leq f_{3}^{}$ are the eigenvalues of
matrix $f$. The inert fermions can serve as the dark matter if and
only if their relic density is consistent with the cosmological
observation.

Before calculating the relic density of the inert fermions, we need
clarify the properties of the gauge and Higgs bosons in the present
model since they are essential to the annihilation of the inert
fermions. There exists a $U(1)_{X}^{}$ gauge field $C_{\mu}^{}$ in
addition to the SM gauge fields $B_{\mu}^{}$ and
$W_{\mu}^{i}\,(i=1,2,3)$. Since the new Higgs doublet with
\textit{vev} carries both $U(1)_{Y}^{}$ and $U(1)_{X}^{}$ charge,
the $U(1)_{X}^{}$ gauge field should mix with the SM ones. By
diagonalizing the vector boson mass matrix, we obtain the charged
bosons $W^{\pm}_{\mu}=\frac{1}{\sqrt{2}}(W_{\mu}^{1}\mp i
W_{\mu}^{2})$ with the mass
$m_{W}^{2}=\frac{1}{2}g^{2}_{}(\langle\phi\rangle^{2}_{}+
\langle\eta\rangle^{2}_{})$, the photon $A_{\mu}^{}=B_{\mu}^{}\cos
\theta + W_{\mu}^{3}\sin\theta $ with the mixing angle $\tan \theta
= g'/g$ as well as the two massive neutral vector bosons
$Z_{\mu}^{}$ and $Z'^{}_{\mu}$,
\begin{eqnarray}
\label{gaugeboson}
Z_{\mu}^{}=Z_{\mu}^{0}\cos\xi-C_{\mu}^{}\sin\xi\,,~~Z'^{}_{\mu}=Z_{\mu}^{0}\sin\xi+C_{\mu}^{}\cos\xi\,
\end{eqnarray}
with the masses,
\begin{eqnarray}
\label{gaugemass1}
m_{Z}^{2}&=&(g^{2}_{}+g'^{2}_{})\{(\langle\sigma\rangle^{2}_{}+\langle\eta\rangle^{2}_{})\sin^{2}_{}\theta
+\frac{1}{4}(\langle\phi\rangle^{2}_{}+\langle\eta\rangle^{2}_{})
\nonumber\\&&-\{[(\langle\sigma\rangle^{2}_{}+\langle\eta\rangle^{2}_{})\sin^{2}_{}\theta
-\frac{1}{4}(\langle\phi\rangle^{2}_{} +\langle\eta\rangle^{2}_{})
]^{2}_{}\nonumber\\
&&+\langle\eta\rangle^{4}_{}\sin^{2}_{}\theta\}^{\frac{1}{2}}_{}\}\,,\\
m_{Z'}^{2}&=&(g^{2}_{}+g'^{2}_{})\{(\langle\sigma\rangle^{2}_{}+\langle\eta\rangle^{2}_{})\sin^{2}_{}\theta
+\frac{1}{4}(\langle\phi\rangle^{2}_{}+\langle\eta\rangle^{2}_{})
\nonumber\\&&+\{[(\langle\sigma\rangle^{2}_{}+\langle\eta\rangle^{2}_{})\sin^{2}_{}\theta
-\frac{1}{4}(\langle\phi\rangle^{2}_{} +\langle\eta\rangle^{2}_{})
]^{2}_{}\nonumber\\
&&+\langle\eta\rangle^{4}_{}\sin^{2}_{}\theta\}^{\frac{1}{2}}_{}\}\,
\end{eqnarray}
and the mixing angle,
\begin{eqnarray}
\label{gaugemixing1} \sin2\xi&=& \langle\eta\rangle^{2}_{}\sin\theta
\{[(\langle\sigma\rangle^{2}_{}+\langle\eta\rangle^{2}_{})\sin^{2}_{}\theta
\nonumber\\
&& -\frac{1}{4}(\langle\phi\rangle^{2}_{}+\langle\eta\rangle^{2}_{})
]^{2}_{}
+\langle\eta\rangle^{4}_{}\sin^{2}_{}\theta\}^{-\frac{1}{2}}_{}\,.
\end{eqnarray}
Here $Z_{\mu}^{0}=-B_{\mu}^{}\sin \theta  + W_{\mu}^{3}\cos\theta $
corresponds to the neutral vector boson of the SM. As shown in Eq.
(\ref{range}), $\langle\eta\rangle$ is much smaller than
$\langle\phi\rangle$ and $\langle\sigma\rangle$, thus we obtain
\begin{eqnarray}
\label{gaugemass2}
m_{Z}^{2}\simeq\frac{1}{2}(g^{2}_{}+g'^{2}_{})\langle\phi\rangle^{2}_{}\,,~~~
m_{Z'}^{2}\simeq2g'^{2}_{}\langle\sigma\rangle^{2}_{}\,.
\end{eqnarray}
Meanwhile, the mixing angle (\ref{gaugemixing1}) is tiny and hence
free of the constraint from the precise measurement. In consequence,
$C_{\mu}^{}$ and $Z_{\mu}^{0}$ can be approximatly identified with
$Z'^{}_{\mu}$ and $Z^{}_{\mu}$, respectively.

Let us subsequently consider the Higgs sector,
\begin{eqnarray}
\label{potential} V(\phi,\sigma,\eta) &=&
-m_{1}^{2}\phi^{\dagger}_{}\phi-m_{2}^{2}\sigma^{\dagger}_{}\sigma+M_{\eta}^{2}\eta^{\dagger}_{}\eta
+\lambda_{1}^{}(\phi^{\dagger}_{}\phi)^{2}_{}\nonumber\\
&&+\lambda_{2}^{}(\sigma^{\dagger}_{}\sigma)^{2}_{}+\lambda_{3}^{}(\eta^{\dagger}_{}\eta)^{2}_{}
+\frac{1}{2}\lambda_{4}^{}(\phi^{\dagger}_{}\phi)(\sigma^{\dagger}_{}\sigma)\nonumber\\
&&+\frac{1}{2}\lambda_{5}^{}(\phi^{\dagger}_{}\phi)(\eta^{\dagger}_{}\eta)
+\frac{1}{2}\lambda_{6}^{}(\sigma^{\dagger}_{}\sigma)(\eta^{\dagger}_{}\eta)
\nonumber\\
&&+\mu \sigma \eta^{\dagger}_{}\phi+\textrm{h.c.}\,.
\end{eqnarray}
Similar to \cite{gh2006}, we can deduce the \textit{vev}s,
$\langle\phi\rangle$, $\langle\sigma\rangle$ and
$\langle\eta\rangle$ by minimizing the above scalar potential. For
$\langle\eta\rangle \ll \langle\phi\rangle,\langle\sigma\rangle$,
the contribution from $\eta$ to $\sigma$ and $\phi$ can be
neglected, we thus have the two neutral bosons,
\begin{eqnarray}
h=\frac{1}{\sqrt{2}}\phi-\langle\phi\rangle\,,~~~h'=\frac{1}{\sqrt{2}}\sigma-\langle\sigma\rangle\,,
\end{eqnarray}
which are the linear combinations of the mass eigenstates $h_{1}^{}$
and $h_{2}^{}$,
\begin{eqnarray}
h_{1}^{}= h\cos\beta- h'\sin\beta\,,~~~ h_{2}^{}=
h\sin\beta+h'\cos\beta\,
\end{eqnarray}
with the masses,
\begin{eqnarray}\label{higgsmass1}
m^{2}_{h_{1}^{}}&=&
\lambda_{1}^{}\langle\phi\rangle^{2}_{}+\lambda_{2}^{}\langle\sigma\rangle^{2}_{}
-[(\lambda_{2}\langle\sigma\rangle^{2}_{}-\lambda_{1}^{}\langle\phi\rangle^{2}_{})^{2}_{}\nonumber\\
&&+4\lambda_{4}^{2}\langle\sigma\rangle^{2}_{}\langle\phi\rangle^{2}_{}]^{\frac{1}{2}}_{}\,,\\
m^{2}_{h_{2}^{}}&=&
\lambda_{1}^{}\langle\phi\rangle^{2}_{}+\lambda_{2}^{}\langle\sigma\rangle^{2}_{}
+[(\lambda_{2}\langle\sigma\rangle^{2}_{}-\lambda_{1}^{}\langle\phi\rangle^{2}_{})^{2}_{}\nonumber\\
&&+4\lambda_{4}^{2}\langle\sigma\rangle^{2}_{}\langle\phi\rangle^{2}_{}]^{\frac{1}{2}}_{}\,
\end{eqnarray}
and the mixing angle,
\begin{eqnarray}
\label{higgsmixing} \tan 2\beta
=\frac{2\lambda_{4}^{}\langle\sigma\rangle\langle\phi\rangle}
{\lambda_{2}^{}\langle\sigma\rangle^{2}_{}-\lambda_{1}^{}\langle\phi\rangle^{2}_{}}\,.
\end{eqnarray}
Similar to \cite{bgm2006}, here the couplings of $h_{1}^{}$ and
$h_{2}^{}$ to the SM gauge bosons, quarks and charged leptons have
the same structure as the corresponding Higgs couplings in the SM,
however, their size is reduced by $\cos\beta$ and $\sin\beta$,
respectively. For $\langle\sigma\rangle \simeq
\mathcal{O}(\textrm{TeV})$, the mixing angle $\beta$ and the mass
splitting between $h_{1}^{}$ and $h_{2}^{}$ may be large. In
consequence, there could be significant impact on the Higgs searches
at LHC \cite{bgc2006}. For example, the couplings of the lighter
$h_{1}^{}$ to the quarks and leptons would even vanish in the
extreme case $\beta =\frac{\pi}{2}$.

\begin{figure} \vspace{4.5cm}
\epsfig{file=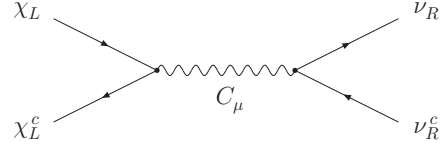, bbllx=6.0cm, bblly=7.0cm, bburx=16cm,
bbury=17cm, width=6.5cm, height=6.5cm, angle=0, clip=0}
\vspace{-8cm} \caption{\label{annihilation} The inert fermions
annihilate into the right-handed neutrinos through the gauge
couplings. }
\end{figure}

We now discuss the possibility of the inert fermions as the dark
matter. The pairs of the inert fermions have the gauge couplings to
$C_{\mu}^{}$ and the Yukawa couplings to $h'$. For
$\langle\eta\rangle\ll\langle\phi\rangle,\,\langle\sigma\rangle$,
$C_{\mu}^{}$ can be looked on as the mass eigenstate $Z'^{}_{\mu}$.
Furthermore, since the systematic analyses of the implication from
the quartic interaction
$\lambda_{4}^{}(\sigma^{\dagger}_{}\sigma)(\phi^\dagger_{}\phi)$ on
the relic density of the inert fermions will be presented elsewhere,
we for simplicity take $\lambda_{4}^{}=0$ and hence $h$ and $h'$ are
exactly identified with $h_{1}^{}$ and $h_{2}^{}$ with the masses,
\begin{eqnarray}
\label{higgsmass2} m^{2}_{h_{1}^{}}=
2\lambda_{1}^{}\langle\phi\rangle^{2}_{}\,,~~~m^{2}_{h_{2}^{}}=
2\lambda_{2}^{}\langle\sigma\rangle^{2}_{}\,,
\end{eqnarray}
respectively. For the purpose of calculating the relic density of
the inert fermions, we take $\langle\sigma\rangle=720\,\textrm{GeV}$
and then give $m_{Z'}^{}\simeq 360\,\textrm{GeV}$, $m_{h'}^{}\simeq
600\,\textrm{GeV}$ with $\lambda_{2}^{}=0.35$,
$m_{\chi_{1,2,3}^{}}^{}\simeq 200\,\textrm{GeV}$ with
$f_{1,2,3}^{}=0.28$ by using Eqs. (\ref{gaugemass2}),
(\ref{higgsmass2}) and (\ref{darkmass2}), respectively. With this
mass spectrum, as shown in Fig. \ref{annihilation}, the channel of a
pair of the inert fermions to a pair of the right-handed neutrinos
should dominate the annihilation process of the inert fermions. We
calculate
\begin{eqnarray}
\label{crosssection}
\sigma_{i}^{}\,v&=&\frac{g'^{4}_{}}{96\,\pi}\,\frac{s\,-\,m_{\chi_{i}^{}}^{2}}
{(s\,-\,m_{Z'}^{2})^{2}_{}\,+\,m_{Z'}^{2}\,\Gamma_{Z'}^{2}}\,,
\end{eqnarray}
where $\sigma_{i}^{}$ is the annihilation cross section of a pair of
$\chi_{i}^{}$ to a pair of $\nu_{R}^{}$, $v$ is the relative speed
between the two $\chi_{i}^{}$'s in their center-of-mass system
(cms), $s$ is the usual Mandelstam variable, and
\begin{eqnarray}
\label{decaywidth} \Gamma_{Z'}^{}&=
&\frac{g'^{2}_{}}{24\,\pi}\,m_{Z'}^{}\,
\end{eqnarray}
is the decay width of $Z'$. Comparing the annihilation rate
$\Gamma_{i}^{}=n_{\chi_{i}^{}}^{eq}\langle\sigma_{i}^{}v\rangle$ to
the Hubble constant, we find that the freeze out should happen at
the temperature $T_{F}^{}\simeq 10\,\textrm{GeV}$. Here
$\langle\sigma_{i}^{}v\rangle\simeq 2.6\,\textrm{pb}$ is the
thermal-average cross section. The relic density of the inert
fermions is then approximately given by
\begin{eqnarray}
\Omega_{\chi}^{}h^{2}_{}\simeq
\sum_{i=1}^{3}\frac{0.1\,\textrm{pb}}{\langle\sigma_{i}^{}v\rangle}\simeq
0.1\,,
\end{eqnarray}
which is equal to the right amount \cite{pdg2006} of the relic
density for the cold dark matter. Thus the inert fermions in the
present model can serve as the candidate for the cold dark matter.

Note that for the above parameters, the right-handed neutrinos will
also decouple at $T_{F}^{}$. So the ratio of the relic density of
the right-handed neutrinos over that of the left-handed neutrinos is
about \cite{kt1990}
\begin{eqnarray}
\label{neutrinodensity}
\frac{n_{\nu_{R}^{}}^{}}{n_{\nu_{L}^{}}^{}}&\simeq&\frac{g_{\ast
S}^{}(\textrm{MeV})}{g_{\ast S}^{}(\textrm{10\,GeV})}= \mathcal{O}(
0.1)\,,
\end{eqnarray}
which is consistent with the current cosmological observation.

It is convenient to extend the present model with certain global
symmetry, after which is spontaneously broken near the Planck scale,
the pNGBs \cite{weiss1987,bhos2005,ghs2007} are expected to arise
and then explain the quintessence dark energy \cite{wetterich1988}.
For example, we replace the Lagrangian (\ref{yukawa1}) and
(\ref{cubic}) by
\begin{eqnarray}
\label{lagrangian} \mathcal{L} &\supset&
-\,\mu_{0}^{}\,\sigma\,\eta_{0}^{\dagger}\, \phi\,-\,\sum_{i\neq
j}^{}\,h_{ij}^{}\xi_{ij}^{}\,\sigma\,\eta_{ij}^{\dagger}\,\phi\,
\nonumber\\
&& -\,y_{ii}^{}\,\overline{\psi_{Li}^{}}\,\eta_{0}^{}\,\nu_{Ri}^{}\,
-\,\sum_{i\neq
j}^{}\,y_{ij}^{}\,\overline{\psi_{Li}^{}}\,\eta_{ij}^{}\,\nu_{Rj}^{}\,\nonumber\\
&&-\,\sum_{i\neq j,k\neq
\ell}^{}\,z_{ij,k\ell}^{}\,\xi_{k\ell}^{\dagger}\,\xi_{ij}^{}\,\eta_{ij}^{\dagger}\,\eta_{k\ell}^{}\,+\,\textrm{h.c.}\,,
\end{eqnarray}
which is supposed to be invariant under a global $U(1)^{3}_{}$
symmetry, generated by the independent phase transformations of
three Higgs singlets, $\xi_{ij}^{}\equiv \xi_{ji}^{\ast}\,(i\neq
j)$, in the limit of vanishing $y_{ij}^{}\,(i\neq j)$. In other
words, the $U(1)^{3}_{}$ is broken down to a $U(1)^{2}_{}$ due to
the presence of $y_{ij}^{}\,(i\neq j)$. Thus after the global
symmetry is broken by the \textit{vev}s,
$\langle\xi_{ij}^{}\rangle\simeq M$, there will be two massless
Nambu-Goldstone bosons (NGBs) and one pNGB, which is associated with
the neutrino mass-generation. Similar to \cite{ghs2007}, a typical
term in the Coleman-Weinberg effective potential of this pNGB has
the form,
\begin{eqnarray}
V(Q) &\simeq& V_{0}^{}\cos (Q/M) \,,
\end{eqnarray}
with $V_{0}^{}=\mathcal{O}(m_{\nu}^{4})$. It is well-known that if
$M$ is near the Planck scale $M_{\textrm{Pl}}^{}$, $Q$ will obtain a
mass of the order of $\mathcal{O}(m_{\nu}^{2}/M_{\textrm{Pl}}^{})$
and can be a consistent candidate for the quintessence dark energy.
Therefore, the intriguing coincidence between the neutrino mass
scale $\sim\, 10^{-2}\,\textrm{eV}$ and the dark energy scale
$\sim\,10^{-3}\,\textrm{eV}$ can be naturally understood. The
leading phenomenology of mass varying neutrinos
\cite{gwz2003,bgwz2003} is very interesting and can be tested in the
present and upcoming experiments \cite{knw2004}.

\begin{figure*} \vspace{3cm}
\epsfig{file=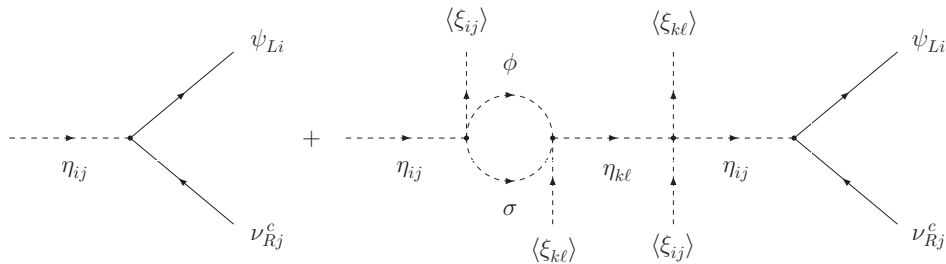, bbllx=6.0cm, bblly=7.0cm, bburx=16cm,
bbury=17cm, width=6.5cm, height=6.5cm, angle=0, clip=0}
\vspace{-5.5cm} \caption{\label{decay} The new Higgs doublets decay
into the leptons at one-loop order. Here $i\neq j $, $k\neq \ell$
and $ij\neq k\ell$. }
\end{figure*}

In the model described by the Lagrangian (\ref{lagrangian}), the
CP-violation and out-of-equilibrium decays of the new Higgs
doublets, as shown in Fig. \ref{decay}, can produce a lepton
asymmetry stored in the left-handed leptons and an equal but
opposite lepton asymmetry stored in the right-handed neutrinos. The
left-handed lepton asymmetry will be partially converted to the
baryon asymmetry through the sphaleron processes \cite{krs1985} and
then explain the matter-antimatter asymmetry of the universe. This
new type of leptogenesis \cite{fy1986} with the conserved lepton
number is called neutrinogenesis \cite{dlrw1999}. For simplicity,
here we will not present the detailed calculation, which is similar
to that in a previous work \cite{ghs2007}.

In this paper, the mass origin at the TeV scale for the Dirac
neutrinos and the dark matter has been successfully unified in a
$U(1)_{X}^{}$ gauge extension of the SM. After the $U(1)_{X}^{}$
breaking, the Dirac neutrinos can obtain small Yukawa couplings to
the SM Higgs and then realize the tiny masses, while the inert
fermions can acquire the masses of a few hundred GeV. The inert
fermions can annihilate to realize the right amount of the relic
density for the cold dark matter. Furthermore, the SM Higgs boson
could no longer be a mass eigenstate, and its signatures at LHC
could be interesting to modify. Finally, in the presence of the
proper global symmetry, after which is spontaneously broken near the
Planck scale, the pNGB associated with the neutrino mass-generation
can provide the consistent candidate for the dark energy, meanwhile,
the matter-antimatter asymmetry \cite{bk1990} of the universe can be
generated via the out-of-equilibrium decays of the heavy Higgs
doublets with the CP-violation.

\textbf{Acknowledgments}: I am greatly indebted to Utpal Sarkar and
Alexei Yu. Smirnov for enlightening discussions and suggestions.

\end{document}